\documentclass[aps,pra,twocolumn,groupedaddress]{revtex4}
\usepackage{graphicx}

\newcommand{\bea}{\begin{eqnarray}}

\newcommand{\eea}{\end{eqnarray}}

\newcommand{\beq}{\begin{equation}}

\newcommand{\eeq}{\end{equation}}

\newcommand{\lav}{\langle}

\newcommand{\rav}{\rangle}

\begin{document}

\def \tr{{\mbox{tr~}}}
\def \ra{{\rightarrow}}
\def \ua{{\uparrow}}
\def \da{{\downarrow}}
\def \be{\begin{equation}}
\def \ee{\end{equation}}
\def \ba{\begin{array}}
\def \ea{\end{array}}
\def \bea{\begin{eqnarray}}
\def \eea{\end{eqnarray}}
\def \nn{\nonumber}
\def \l{\left}
\def \r{\right}
\def \half{{1\over 2}}
\def \etal{{\it {et al}}}
\def \cH{{\cal{H}}}
\def \cM{{\cal{M}}}
\def \cN{{\cal{N}}}
\def \cQ{{\cal Q}}
\def \cI{{\cal I}}
\def \cV{{\cal V}}
\def \cG{{\cal G}}
\def \cF{{\cal F}}
\def \cZ{{\cal Z}}
\def \bS{{\bf S}}
\def \bI{{\bf I}}
\def \bL{{\bf L}}
\def \bG{{\bf G}}
\def \bQ{{\bf Q}}
\def \bK{{\bf K}}
\def \bR{{\bf R}}
\def \br{{\bf r}}
\def \bu{{\bf u}}
\def \bq{{\bf q}}
\def \bk{{\bf k}}
\def \bp{{\bf p}}
\def \bz{{\bf z}}
\def \bx{{\bf x}}
\def \bpsi{{\bar{\psi}}}
\def \tJ{{\tilde{J}}}
\def \W{{\Omega}}
\def \e{{\epsilon}}
\def \lam{{\lambda}}
\def \L{{\Lambda}}
\def \a{{\alpha}}
\def \t{{\theta}}
\def \b{{\beta}}
\def \g{{\gamma}}
\def \D{{\Delta}}
\def \d{{\delta}}
\def \w{{\omega}}
\def \s{{\sigma}}
\def \f{{\varphi}}
\def \x{{\chi}}
\def \e{{\epsilon}}
\def \h{{\eta}}
\def \G{{\Gamma}}
\def \z{{\zeta}}
\def \hatt{{\hat{\t}}}
\def \hn{{\bar{n}}}
\def \vk{{\bf{k}}}
\def \vq{{\bf{q}}}
\def \gk{{\g_{\vk}}}
\def \nd{{^{\vphantom{\dagger}}}}
\def \yd{^\dagger}
\def \av#1{{\langle#1\rangle}}
\def \ket#1{{\,|\,#1\,\rangle\,}}
\def \bra#1{{\,\langle\,#1\,|\,}}
\def \braket#1#2{{\,\langle\,#1\,|\,#2\,\rangle\,}}


\title{Noise Correlations in low-dimensional systems of ultracold atoms}

\author{L. Mathey$^1$, A. Vishwanath$^{2,3}$ and  E. Altman$^4$}
\affiliation{$^1$Joint Quantum Institute, National Institute of Standards and Technology,
and University of Maryland, Gaithersburg, Maryland 20899, USA \\
$^2$Department of Physics, University of California, Berkeley, CA 94720 \\
$^3$Materials Science Division, Lawrence Berkeley National Laboratories, Berkeley, CA 94720 \\
$^4$Department of Condensed Matter Physics, The Weizmann Institute of Science Rehovot, 76100, Israel}

\date{\today}

\begin{abstract}
We derive relations between standard order parameter correlations and the noise correlations in time of flight images, which are valid for systems with long range order as well as low dimensional systems with algebraic decay of correlations.
Both Bosonic and Fermionic systems are considered. For one dimensional Fermi systems we show that the noise correlations are equally sensitive to spin, charge and pairing correlations and may be used to distinguish between fluctuations in the different channels. This is in contrast to linear response experiments, such as Bragg spectroscopy, which are only sensitive to fluctuations in the particle-hole channel (spin or charge). For Bosonic systems we
find a sharp peak in the noise correlation at opposite momenta that signals pairing correlations in the depletion cloud. In a condensate with true long range order, this peak is a delta function and we can use Bogoliubov theory to study its temperature dependence. Interestingly we find that it is enhanced with temperature in the low temperature limit. In one dimensional condensates with only quasi-long range (i.e. power-law) order the peak in the noise correlations also broadens to a power-law singularity. 
\end{abstract}


\maketitle

\section{Introduction}
The ability to trap ultracold atoms in tightly confined tubes formed
by an optical lattice has opened the door for the controlled study of
one dimensional physics. Such systems have been used to investigate
ground state correlations \cite{paredes,weiss} as well as dynamics
\cite{stoeferle} and transport \cite{fertig} of strongly interacting
bosons in one dimension. One dimensional traps of ultracold fermions
have also been realized. The ability to control the interactions has
been demonstrated using s-wave Feshbach resonances for fermions with
spin \cite{moritz}, and with p-wave resonances for spin-polarized
fermions \cite{Gunter}.

From the theoretical viewpoint, one dimensional systems provide good
starting points to study strong correlation physics. Because of the
enhanced quantum fluctuations, continuous symmetries cannot be broken
in generic one dimensional systems, and mean field theories
fail. Nevertheless the Luttinger liquid framework provides a well
developed formalism to treat these systems theoretically (see for
example \cite{giamarchi_book}). In place of a mean field order
parameter the tendencies to ordering manifest themselves by slowly
decaying algebraic correlations and correspondingly divergent
susceptibilities. A diverging susceptibility in a particular channel
implies that coupling an array of tubes in the transverse direction
would lead to true order in that channel. Thus weakly coupled one
dimensional systems provide a possible theoretical route to
investigate open questions regarding competing orders in higher
dimensions.

But the absence of long range order, which makes one dimensional
systems interesting also complicates the ways by which these systems
can be probed.  Various methods have been proposed and used to probe
specific correlations. For example time of flight \cite{paredes} as
well as interference experiments \cite{interference} can probe single
particle correlations. Bragg \cite{DynStructFact, StatStructFact} and
lattice modulation \cite{stoeferle} spectroscopies can measure
dynamic density correlations . In \cite{noise_1d_short} we proposed
that noise correlations can be used as a highly flexible probe of 1D
Fermi systems. In particular this method is sensitive to a wide range
of correlations including spin, charge and pairing, and it treats the
various channels on an equal footing. These results, obtained using the effective
Luttinger liquid theory were confirmed by Luscher et. al. using a numerical
simulation of a microscopic model.

In this paper we provide the
detailed theory of quantum noise interferometry in one dimensional
Fermi systems and extend it to interacting Bose systems.

Atomic shot noise in time of flight imaging was proposed in Ref.
\cite{ehud} as a probe of many body correlations in systems of ultra
cold atoms. Specifically what is measured using this approach is the
momentum space correlation function of the atoms in the trap
\be
\mathcal{G}_{\a\a'}(k,k')=\av{n_{\a k} n_{\a'k'}}-\av{n_{\a
    k}}\av{n_{\a' k'}}, \label{G}
\ee
where $\a,\a'$ are spin indices,
$k, k'$ are momenta, and $n_{\a, k}$ and $n_{\a', k'}$ are the
occupation operators of the corresponding momentum and internal state.
This relies on the assumption that the atoms are approximately
non-interacting from the time they are released from the trap (even if
they were rather strongly interacting in the trap).  Here we envision
a system of one dimensional tubes, that allows expansion only in the
axial direction. To ensure weak interactions during time of flight,
the radial confinement in the tubes can be reduced simultaneously with
release of the atoms from the global trap.

Recent experiments demonstrated the ability to detect many-body
correlations by analysis of the noise correlations (\ref{G}). For
example long ranged density-wave correlations (induced by external
lattice) were seen in a system of ultracold bosons\cite{BlochHBT,PortoNoise} and
fermions\cite{BlochFermions} in deep optical lattices.  The sign of
the correlation, peak or dip, depended on the statistics as expected,
demonstrating that the effect is essentially a generalization of the
Hanbury-Brown Twiss effect.  Pairing correlations in fermions
originating from a dissociated molecular condensate were also observed
in experiment\cite{markus}. All of these demonstrations involved
states with true long range order that are easily related to simple
two particle effects such as Hanbury-Brown Twiss or bound state
formation. However, one of the main points of Ref. \cite{ehud} was
that the noise correlations can be used to measure more general
correlation functions in many-body systems, even in the absence of
true long range order (LRO).  Here we demonstrate this idea by
developing a detailed theory for the noise correlations in one
dimensional systems of bosons and fermions.

The analysis is laid out in the following order. In section \ref{SL_F}
we consider a model system of spinless fermions in one dimension.  To
elucidate the connection between noise correlation and ordering
tendencies we first assume in section \ref{subsec:LRO} the presence of
true long range order. This allows to provide a clear physical picture
for the way order in spin, charge (particle-hole) or pairing
(particle-particle) channels translate into sharp features (in this
case delta function peaks) in the noise correlations. In section
\ref{SLFQO} we shall consider the actual system of interacting
fermions in one dimension where only power law order parameter
correlations exist (quasi-long range order).  This system is
characterized by a tendency to charge density wave (CDW) ordering for
repulsive interactions, and to superconducting order (SC) for
attractive interactions. The naive expectation is that the delta
function peaks will be replaced by power-law singularities with a
power that reflect the algebraic order parameter correlations.
Interestingly we find that such a simple relation exists only if there
are order parameter correlations that decay with a sufficiently slow
power. This is the case beyond a critical interaction strength
(repulsive or attractive).  At weak interactions, on the other hand,
the situation is more subtle and singular signatures of both the
dominant and sub-dominant ordering tendencies are seen in the noise
correlations (see Fig. \ref{noise_SL}b,c).  We compare these results
with the information that can be extracted from measuring the static
structure factors in the spin and charge channels.

\begin{figure}
\includegraphics[width=8cm]{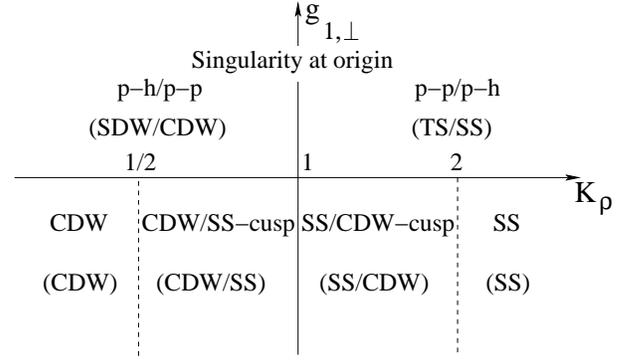}
\caption{\label{PD_noise_S} "Phase Diagram" summarizing the singular signatures in the noise of the various order parameter correlations in a one dimensional spinfull Fermi system. $g_{1\perp}$ is a backscattering parameter and $K_\rho$ the luttinger parameter in the charge sector. The divergent susceptibilities in the different regimes are in brackets. For positive backscattering only weak signatures of order parameter correlations are seen through a weak singularity at the origin. On the other hand in the spin gapped phase at negative backscattering the noise correlations give detailed information on both the CDW and SSC correlations.}
\end{figure}

In section \ref{spin_section}, we move on to analyze the noise
correlations in one dimensional systems of interacting spin-$1/2$
fermions. In the long wavelength effective field theory the
interaction is parameterized by two Luttinger parameters $K_\rho$ and
$K_\s$ corresponding to the spin and charge sectors, as well as a
backscattering parameter $g_{1\perp}$. The ordering tendencies of this
system are summarized in the phase-diagram in Fig. \ref{PD_noise_S}
(See \cite{giamarchi}). We find that the noise correlations provide direct
information on the real space order parameter correlations only for
negative backscattering.  In this case a spin-gap opens and the system
is characterized by competing CDW and SSC correlations, which are
revealed by the noise correlations.

In section \ref{B}, we turn to interacting Bose systems. We first study the noise correlation within Bogoliubov theory for a condensate with true long range order (section \ref{BEC}). The interesting feature her is a peak in the noise correlations (\ref{G}) at $k+k'=0$. This can be
simply understood as the signature of pairing correlations in the
Bogoliubov wave-function, describing quantum depletion of the
condensate.  We compute the evolution of this signature with
increasing temperature. In addition we point out the existance of sharp dips in
the noise correlations along the lines $k'=0$ and $k=0$, these reflect
correlations between the condensate and the quantum depletion cloud.
In section \ref{QC} we move on to discuss interacting one
dimensional Bose liquids, where the single particle density matrix
decays as a power-law with distance. We show how the sharp features in
the noise correlation function on the lines $k+k'=0$, $k=0$, and
$k'=0$ broaden to power-law singularities.

\section{\label{SL_F}Spinless Fermions}
A system of interacting spinless fermions is perhaps the simplest
Fermi system with a non trivial competition of ordering
tendencies. Such systems can be implemented in experiments using fully
polarized fermionic atoms that interact via an odd angular momentum
channel.  The interaction strength can be tuned by using a $p$-wave
Feshbach resonance \cite{Gunter}.  Another possibility is to use a
Bose-Fermi mixture, where the phonons of the bosonic superfluid
mediate effective interactions between the fermions \cite{mathey}.

In the restricted geometry of 1D the Fermi surface consists only of
two points, the left (L) Fermi point at $-k_f$ and the right (R) Fermi
point at $+k_f$.  We introduce the left- and right-moving fields
$\psi_L$ and $\psi_R$ through:
\bea
\psi(x) & = & e^{-i k_f x}\psi_L + e^{i k_f x}\psi_R
\label{split}
\eea
$\psi_L$ and $\psi_R$ are slowly varying fields, because the rapidly
oscillating phase factors $e^{\pm i k_f x}$ have been separated out.
The natural order parameters that characterize this system are the
charge density and superconducting (pairing) operators:
\bea
O_{CDW} & =  & \psi_R^\dagger\psi_L=\rho_{2k_f}\\
O_{SC} & = & \psi_R \psi_L
\eea
The central problem we address in this section concerns the connection
between the correlations of these physical order parameters and the
observable noise correlation signal.

\subsection{Long range order}\label{subsec:LRO}
The relation between order parameter correlations and sharp features
in the noise correlation function is most apparent when the system
supports true long range order. Of course one dimensional Fermi
systems generically do not display spontaneous long range order in
either the CDW or SC channels. However long range order can be induced
by an external potential (e.g. a superlattice potential) or it can
form spontaneously in a weakly coupled array of one dimensional
systems.  Moreover the intuition gained from the exercise will be
useful in approaching the more interesting case of power-law order
parameter correlations (quasi long range order), which will be
considered in the following sections.

Consider first the case of long range pairing correlations (i.e. SC
order). Take $k$ ($k'$) to be near the right (left) Fermi points, and
$q\equiv k-k_F$, $q'\equiv k'+k_F$ deviations from those Fermi
points. It is now useful to write the noise correlations explicitly in
terms of the pairing correlations
\begin{widetext}
\bea
\av{n_{k} n_{k'}}&\approx& {1\over L^2}\int dX dX'dr dr'e^{i(q+q')(X-X')}e^{i(q-q')(r-r')/2}
\av{\psi\yd_R(X+r/2)\psi\yd_L(X-r/2)\psi\nd_L(X'-r'/2)\psi\nd_R(X'+r'/2)} \nn\\
&=& {1\over L^2}\int dX dX'dr dr'e^{i(q+q')(X-X')}e^{i(q-q')(r-r')/2}
\av{B\yd_r(X)B\nd_{r'}(X')}
\label{NoiseLRO}
\eea
\end{widetext}
Here the operator $B_r(X)$ creates a fermion pair whose constituents
are separated by $r$ and their center of mass coordinate is $X$. Note
that we have dropped terms that undergo rapid $2k_F$ spatial
oscillations and therefore vanish under integration.  The existence of
long range pairing correlations implies
\be
\av{B\yd_r(X)B\nd_{r'}(X')}\to
\Phi(r)\Phi(r')
\ee
at long distances ($|X-X'|\to\infty$).  Here
$\Phi(r)=\av{\psi\yd(X+r/2)\psi\yd(X-r/2)}$ is the translationally
invariant pairing wave function. The long range saturation of the
correlation function leads to a singular contribution to
(\ref{NoiseLRO}) at $q=-q'$:
\be
\av{n_{k} n_{-k'}}\sim \d(q+q')|{\tilde \Phi}(q)|^2
\ee
We thus conclude that the noise correlations in this case are directly
related to the long distance limit of the pairing correlations.  This
is one way to formally justify the mean-field decoupling of the noise
correlation function carried out in Ref. [\onlinecite{ehud}], which
gives:
\be\label{contract1}
\av{n_\bk n_{\bk'}}-\av{n_\bk n_{\bk'}}=|\av{\psi_\bk\psi_{-\bk}}|^2\d(\bk+\bk')
\ee

A very similar analysis follows for the case of long range order in
the particle-hole cannel. Take for example CDW order at the
wave-vector $2k_F$. In this case we should write the noise correlation
function in terms of the density correlations to expose their singular
contribution:
\begin{widetext}
\bea
\av{n_{k} n_{k'}}&=& -{1\over L^2}\int dX dX'dr dr'e^{i(q-q')(X-X')}e^{i(q+q')(r-r')/2}
\av{\psi\yd_R(X+r/2)\psi\nd_L(X-r/2)\psi\yd_L(X'-r'/2)\psi\nd_R(X'+r'/2)} \nn\\
&=& -{1\over L^2}\int dX dX'dr dr'e^{i(q+q')(X-X')}e^{i(q-q')(r-r')/2}
\av{\rho_{2k_F,r}(X)\rho_{2k_F,r'}(X')}
\label{NoiseLROCDW}
\eea
\end{widetext}
Note the minus sign in front of the integral, which resulted from
commuting the Fermi operators to obtain an expression written in terms
of density wave correlations. Long range CDW correlations imply
\be
\av{\rho_{2k_F,r}(X)\rho_{2k_F,r'}(X')}\to \Xi(r)\Xi(r').
\ee
which leads to a singular contribution to (\ref{NoiseLROCDW}) for $q=q'$
\be
\av{n_k n_{k'}}\sim -\d(q-q')|{\tilde \Xi}(q)|^2
\ee
This justifies the corresponding mean field decoupling of the noise
correlation function:
\be\label{contract2}
\av{n_k n_{k'}}-\av{n_k n_{k'}}=~-|\av{\psi\yd_{k+2k_F}\psi\nd_{k}}|^2\d(k-k')
\ee
Again, as in the SC case, we see that the noise correlations are
directly related to the long distance limit of the non decaying order
parameter correlation.

\begin{figure}
\includegraphics[width=5cm]{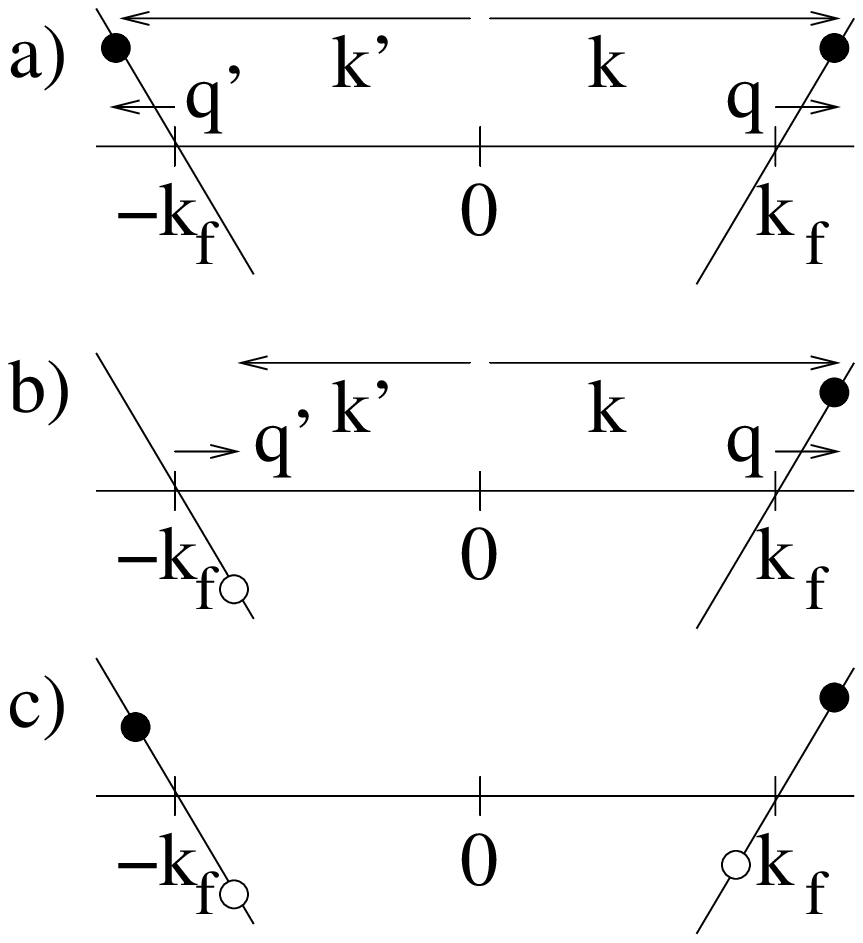}
\caption{\label{fluct} Schematic representation of several types of
  fluctuations.  In these diagrams, $k$ and $k'$ represent the momenta
  of the atoms and holes, and $q$ and $q'$ the momenta relative to the
  Fermi points.  (a) A pairing fluctuation (or Cooper pair), 
  which is the dominant
  fluctuation in the SC phase. These fluctuations result in positive
  noise correlations along $q=-q'$.  (b) A particle-hole (p-h) fluctuation 
  associated with a CDW state. This fluctuation results in
  negative correlations along $q=q'$.  (c) Two p-h
  pairs. This fluctuation results in both positive and negative correlations.
  Positive correlations for $q=-q'$ and negative for $q=q'$.}
\end{figure}

We note two obvious distinctions between the cases of order in the
particle-particle (pairing) and particle-hole channels. In the former
the singular correlations are between particles with opposite momenta
and are positive correlations. In the latter case, by contrast, the
singular correlations are between particles whose momenta differ by
$2k_F$ and are negative, that is anti-correlations. To better
understand the origin of this effect it is worthwhile to inspect the
mean field wave functions, which sustain the respective broken
symmetries. For the superconducting order the mean field wave-function
is the BCS state representing a pair condensate:
\be
\ket{\Psi_{SC}}=\prod_k\left[u_k+v_k\psi\yd_k\psi\yd_{-k}\right]\ket{0}
\label{BCSWF}
\ee
Contrary to a filled Fermi sea the particle number $n_k$ in a specific
$k$ point is not definite in this wave function. However, if a
particle is found at $k$, there is with certainty another one at
$-k$. This implies positive correlation between $n_k$ and $n_{-k}$ as
visualized in Fig. \ref{fluct} (a).

The CDW state on the other hand may be viewed as a condensate of
particle-hole pairs on top of the filled Fermi sea. It is then written
in a way that exposes the similarity to the BCS state:
\be
\ket{\Psi_{CDW}}=\prod_k\left[u_k+v_k\psi\yd_{k+2k_F}\psi\nd_{k}\right]\ket{FS}
\ee
Here $\ket{FS}$ denotes a full fermi sea wavefunction. As in the BCS state, the particle number $n_k$ at a specific $k$ point is not definite. Now however, if one finds a particle at $k>k_F$ there will
be with certainty a missing particle (hole) in the Fermi sea at
$k-2k_F$. This implies anti-correlation between $n_k$ and
$n_{k-2k_F}$ as visualized in Fig. \ref{fluct} (b).

\subsection{\label{SLFQO} Quasi-order}
Because of strong quantum fluctuations, an actual one dimensional
Fermi system cannot sustain true long range order such as the SC and
CDW orders discussed above. Instead, at zero temperature a critical
phase with power law correlations, or quasi long range order, is
established. If the power-law decay is sufficiently slow then it is
reasonable to expect that it would still make a singular contribution
to the integrals in (\ref{NoiseLRO}) or (\ref{NoiseLROCDW}). To
calculate the resulting singularity and its dependence on the system
parameters we use the low energy Luttinger liquid theory.  The basic
idea is that the asymptotic low energy and long wave-length properties
of the interacting one dimensional Fermi system are captured by a
universal harmonic theory:
\bea\label{S_LL}
S &= & \frac{1}{2 \pi K}\int dx d\tau \left[c(\partial_x\Theta)^2+c^{-1}(\partial_\tau\Theta)^2 \right]\nn\\
&=&  \frac{K}{2 \pi}\int dx d\tau \left[c(\partial_x\Phi)^2+c^{-1}(\partial_\tau\Phi)^2 \right]
\eea
Here $K$ is the Luttinger parameter, which determines the power-law
decay of long range correlations. $c$ is a sound velocity, which we
will henceforth set to 1. The bosonization identity, which relates the
bosonic fields to the fermion operators is:
\bea
\psi_{R/L}(x) & = & \frac{1}{\sqrt{2 \pi \alpha}}
e^{\pm i\Theta(x)+i \Phi(x)},
\label{bosonization}
\eea
with the commutation relation:
\bea
[\Theta(x), \Phi(0)] & = &
\frac{1}{2} \log\frac{\alpha + i x}{\alpha - i x}. \label{com}
\eea
$\a$ is an artificial cutoff of the bosonization procedure which must
be sent to zero at the end of the calculation. Implicit in the action
(\ref{S_LL}) is also a physical short distance cutoff $x_0$, taken in
most cases to be of order $1/k_F$.  Here $\Phi$ is the phase field of
the Fermi operator.  $\Pi={1\over \pi}\nabla\Theta$ is the smooth
($k\approx 0$) component of the density fluctuation, and by
(\ref{com}), is also canonically conjugate to $\Phi$.

The effective action (\ref{S_LL}) is a free theory in terms of the
bosonic fields. It therefore allows to calculate the needed
correlation functions using simple gaussian quadrature. The fermion
interactions affect the correlation functions only through the
Luttinger parameter $K$. For non interacting fermions $K=1$, $K<1$ for
repulsive interactions and $K>1$ for attractive interactions. In
general $K$ deviates more from $K=1$ the stronger the
interactions. However it is in general not possible to make a precise
connection between the microscopic parameters and the Luttinger
parameter.  Among other things, our analysis points to a way of
extracting this parameter from experiments. 

\begin{figure}[t]
\includegraphics[width=8.5cm]{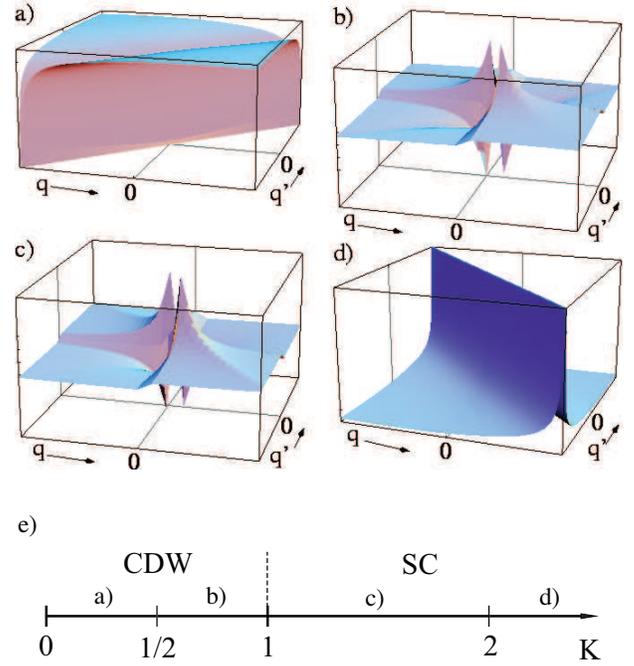}
\caption{\label{noise_SL} 
a) -- d) 
 Noise correlations $\lav n_{q} n_{q'} \rav -
  \lav n_{q} \rav \lav n_{q'} \rav$ of a 1D Fermi system for different
  values of the Luttinger parameter $K$. For a) -- d): $K=0.4$, $0.8$,
  $1.25$ and $2.5$.  For $K<1/2$, we find a negative algebraic
  divergence along $q=q'$, reflecting a quasi-condensate of p-h pairs.
  For $K>2$, we find a positive algebraic divergence along $q=-q'$,
  indicating a quasi-condensate of p-p pairs.  In the intermediate
  regime we see signatures of both p-h and p-p correlations, which are
  singular at the origin. 
 e) `Phase diagram' of a Luttinger liquid of spinless fermions as defined by diverging susceptibilities  
and signature of the order parameter correlations in the noise. For strong interaction, $K<1/2$ and $K>2$, we find diverging noise correlation along a line. For $K<1/2$ it is a negative peak on the line $q=q'$ signaling a particle-hole correlation. For $K>2$ it is a positive peak at $q=-q'$ signaling a particle-particle correlation. In the intermediate regime ($1/2<K<2$) there are signatures of both the leading and sub-leading fluctuations in the noise correlations which diverge at $q=q'=0$. }
\end{figure}

For any value of $K$ the system shows either CDW or SC quasi-long
range order, which means that the correlation functions decay slow
enough to give a divergent susceptibility.  The calculation of these
correlation functions has been given in numerous places (see
e.g. \cite{review, giamarchi_book}).  For the long distance behavior
of $O_{CDW}$ we have $\lav O_{CDW}(x,\tau)O_{CDW}(0,0)\rav \sim \cos(2
k_f) (x^2+c^2\tau^2)^{-(2-\alpha_{CDW})/2}$, and for $O_{SC}$ we have
$\lav O_{SC}(x,\tau)O_{SC}(0,0)\rav \sim
(x^2+c^2\tau^2)^{-(2-\alpha_{SC})/2}$ where the scaling exponents for
CDW and SC are
\bea
\alpha_{CDW} & = & 2-2K\label{chiCDW}\\
\alpha_{SC} & = & 2-2K^{-1}.\label{chiSC}
\eea
The susceptibilities correspond to the spatial and temporal Fourier
transform of the correlation functions, $\chi(k,\omega)\sim\int dx
d\tau e^{ikx+i\omega\tau}\lav O(x,\tau)O(0,0)\rav$.  These will be
divergent at large distances exactly if the scaling exponent of the
operator $O(x,\tau)$ is positive.  As we can see from (\ref{chiCDW})
and (\ref{chiSC}), $\chi_{CDW}$ diverges at $k=2k_f$ for $K<1$, and
$\chi_{SC}$ diverges at $k=0$ for $K>1$.  In this sense of QLRO we say
that the system is in the CDW regime for $K<1$, and in the pairing
regime for $K>1$ \cite{review}, as depicted in Fig. \ref{noise_SL}
e).  We note that from Eq. \ref{S_LL} and \ref{bosonization} one can
read off the duality mapping: $\theta\leftrightarrow\Phi$,
$K\leftrightarrow K^{-1}$, which leaves the action invariant, and maps
the CDW regime $0<K<1$ onto the SC regime, $1<K<\infty$.

%

%
%

With this formalism we derive the noise correlations
$\mathcal{G}(q,q')$, which we show in detail in App. \ref{noisederiv}.
We find
\bea
\mathcal{G}(q, q') & = & \int\frac{e^{iqx_{12}+iq'x_{34}}}{(2\pi)^2 L} \mathcal{F}(x_{12})\mathcal{F}^*(x_{34})(\mathcal{A} -1)
\label{CorrInt}
\eea
with
\bea
\mathcal{F}(x) & = & \Big(\frac{x_0^2}{x_0^2+x^2}\Big)^g\frac{1}{\alpha - i x}
\eea
and
\bea\label{Ah}
\mathcal{A} & = & \Big(\frac{(x_0^2+x_{14}^2)(x_0^2+x_{23}^2)}{(x_0^2 + x_{13}^2)(x_0^2 + x_{24}^2)}\Big)^h.
\eea
The integration in Eq. \ref{CorrInt} is over the three spatial
variables $x_{12}$, $x_{23}$ and $x_{34}$.  The exponents $g$ and $h$
are given by $g = (K+K^{-1}-2)/4$ and $h=(K-K^{-1})/4$.
%

%
%

Following the discussion of the MFA, we note that this integral
'contains' the equal-time correlations of $O_{CDW}$ and $O_{SC}$.
This can be seen by setting $x_{14}=0$ and $x_{23}=0$, which gives the
Fourier transform in $q-q'$ of $-\lav O_{CDW}(1)O_{CDW}(2)\rav$.  If
we set $x_{13}=0$ and $x_{24}=0$, we obtain the Fourier transform of
$\lav O_{SC}(1)O_{SC}(2)\rav$ in $q+q'$.  To discuss this further we
introduce the following variables: $z=(x_{12}-x_{34})/2$,
$h_+=(x_{14}+x_{23})/2$, and $h_-=(x_{14}-x_{23})/2$.  With this,
$\lav n_{q}n_{q'}\rav$, is of the form
\bea\label{other_v}
\lav n_{q}n_{q'}\rav
& \sim &
\int\frac{e^{i (q-q')z + i(q+q')h_-}}{L}
 \mathcal{F}(z+h_-)\mathcal{F}^*(z-h_-)\nonumber\\
& &
\Big(\frac{\Lambda^2 +(h_++h_-)^2}{\Lambda^2+(z+h_+)^2}
\frac{\Lambda^2+(h_+-h_-)^2}{\Lambda^2+(z-h_+)^2}\Big)^h.
\eea
In the limit $K\ll 1$, we have $h\ll 0$.  This enforces the integrand
to be negligible except in regions with $z,h_-\approx 0$.  With this,
the expression approximately evaluates to
\bea\label{scaling1}
\lav n_{q}n_{q'}\rav
& \sim & \mathrm{sgn}(2K-1)|q-q'|^{2K-1}
\eea
In the dual limit, $K\gg 1$, we have $h\gg 0$, which enforces
$z,h_+\approx 0$.  In this limit the integral approximately evaluates
to
\bea\label{scaling2}
\lav n_{q}n_{q'}\rav
 & \sim & \mathrm{sgn}(2K^{-1}-1)|q+q'|^{2K^{-1}-1},
\eea
which can also be inferred from duality.  These contributions are
divergent for $K<1/2$ and $K>2$, and turn out to be the dominant
contribution in theses regimes.

To confirm this expectation, and in order to study the regime
$1/2<K<2$, we evaluate this integral numerically for different values
of $K$.  In Fig. \ref{noise_SL} we show $\mathcal{G}(q, q')$ for $K=
0.4$, $0.8$, $1.25$ and $2.5$. The $K=0.4$ example shows indeed a
power-law divergence of the particle-hole type, which we find
througout the $K<1/2$ regime. For $K>2$ we find a result similar to
the $K=2.5$ example, an algebraic divergence of the particle-particle
type. These two regimes are indicated in Fig. \ref{noise_SL} e).

In between these two regimes, for $1/2<K<2$, we find a regime in which
both p-h as well as p-p correlations exist, i.e. we find precursors of
the near-by competing order.  A simple argument for the qualitative
shape of the noise correlations function in this regime is the
following: If we consider an interacting 1D Fermi gas and treat the
interaction perturbatively, the lowest order contribution would
consist of states that contain two p-h pairs, as two fermions have
been taken from the Fermi sea and put into the unoccupied states above
the Fermi sea (see Fig. \ref{fluct} c)).  Such a state exhibits
qualitatively the noise correlations that are observed in
Fig. \ref{noise_SL} b) and c): positive correlations for $q<0,q'>0$
and $q>0,q'<0$, and negative correlations for $q>0,q'>0$ and
$q<0,q'<0$. To quantify how this affects the line-shape of the noise correlations
we expand $\mathcal{G}(q,q')$ to second order in
$h$ (see App. \ref{intermediate} for details). The result is
\bea\label{NC_int}
\mathcal{G}(q,q') & \sim & -h^2\mathrm{sgn}(q)\mathrm{sgn}(q')\min\Big(\frac{1}{|q|},\frac{1}{|q'|}\Big).
\eea
This expression shows a divergence at $q=q'=0$, and both
particle-particle and particle-hole fluctuations.  Higher order terms
either enhance particle-particle (for $K>1$) or particle-hole ($K<1$)
fluctuations.

Before concluding the section, we compare the static and dynamic
structure factor with the noise correlations discussed in this
section.  The dynamic structure factor can be measured by a stimulated
two-photon process, as has been demonstrated in \cite{DynStructFact}.
It is defined as
\bea
S(k,\omega) & \sim & \lav\rho(k,\omega)\rho(-k,-\omega) \rav.
\eea
 As discussed in Ref. \cite{giamarchi_book}, it is given by
\bea
S(k,\omega) & \sim & K |k| \delta(\omega - c |k|)
\eea
for small $k\approx 0$. For $k\approx 2 nk_f$, it behaves as
\bea\label{DynStructFact}
S(2 n k_f + q, \omega) & \sim & (\omega^2 - c^2 q^2)^{n^2 K-1}
\eea
where $n\neq 0$.  So at $K=1$, coming from larger values, two
additional peaks appear at $\pm 2 k_f$, because the exponent $K-1$ in
Eq. (\ref{DynStructFact}) switches sign.  At $K=1/4$ two additional
peaks appear at $\pm 4 k_f$, and so on.  So the measurement of the
dynamic structure factor, being the Fourier transform of the
density-density correlation function, certainly allows insight into
the CDW phase, but contains no information about the SC phase.  In
contrast, as demonstrated in Fig. \ref{noise_SL}, the measurement of
noise correlations proposed in this paper allows the identification of
pairing and CDW ordering in a single approach and on equal footing,
because both the CDW and the SC correlations are contained in the
expression for the noise correlations.

The static structure factor of a BEC has been measured in
[\onlinecite{StatStructFact}].  It corresponds to the instantaneous
density-density correlations:
\bea
S(k) &\sim& \lav\rho_{-k}\rho_k\rav
\eea
For $k\approx 0$, we have
\bea
S(k\approx 0) & \sim & K |k|
\eea
and for $k\approx 2 n k_f$:
\bea
S(2nk_f +q) & \sim & |q|^{2 n^2 K-1}.
\eea
Here the first set of peaks appears at $K=1/2$, the next set of peaks
at $K=1/8$ and so on.  This is the power-law divergence that dominates
the noise correlations for $K<1/2$, as discussed in the previous
section.  Again there is no signature of SC.

In summary we have derived the noise correlations of a spinless
fermionic LL in this section, and compared it to a MFA result.  We
found different subregimes with qualitatively different behavior in
each of the quasi-phases, summarized in Fig. \ref{noise_SL}, and
discussed how phenomena such as QLRO and competing phases are
reflected in the noise correlations.

\section{\label{spin_section}Spin-$1/2$ Fermions}
In this section we discuss the noise correlations of an
SU(2)-symmetric Fermi mixture \cite{recati}.

Our analysis applies to systems of the form
\bea\label{Hub}
H & = & -t \sum_{\lav ij\rav,\sigma} \psi_{i, \sigma}^\dagger\psi_{j,\sigma} + U \sum_{i,\sigma} n_{i,\uparrow}n_{i,\downarrow}
\eea
i.e. the 1D Hubbard model, or a mixture in a 1D continuum with a
contact interaction:
\bea\label{MixF}
H & = & \sum_\sigma \int \psi_\sigma^\dagger (\frac{-\nabla^2}{2 m})\psi_\sigma
+ U\int n_\uparrow(x)n_\downarrow(x)
\eea
which can both be realized in experiment.
\subsection{\label{SFLRO} Long range order}
Before we turn to the LL picture, we introduce the types of order that
occur in this system, and discuss what kind of signature can be
expected if they develop long-range order, with similar arguments as
in Sect. \ref{subsec:LRO}.

As for the spinless fermions we split the spinful fermionic operators
into left- and right-moving fields:
\bea
\psi_{\uparrow/\downarrow}(x) & = & e^{-i k_f x}\psi_{L,\uparrow/\downarrow} + e^{i k_f x}\psi_{R,\uparrow/\downarrow}.
\label{split_spin}
\eea
We introduce these order parameters \cite{review, giamarchi_book}:
\bea
O_{SS} & = & \psi_{R,\uparrow} \psi_{L,\downarrow}\label{op_ss}\\
O_{TS} & = & \psi_{R,\uparrow} \psi_{L,\uparrow}\label{op_ts}\\
O_{SDW} & = & \psi_{R,\uparrow}^\dagger \psi_{L,\downarrow}\label{op_sdw}\\
O_{CDW} & = & \psi_{R,\uparrow}^\dagger \psi_{L,\uparrow}.\label{op_cdw}
\eea
$O_{SS}$ describes singlet pairing, and therefore contains $\uparrow$
and $\downarrow$ operators, whereas $O_{TS}$ describes the $x$ and $y$
component of triplet pairing, and therefore describes pairing between
equal spin states.  $O_{SDW}$ is the order parameter of the $x$ and
$y$ component of the spin density wave order.

Because these order parameters contain both R and L, as well as
$\uparrow$ and $\downarrow$ operators, we expect correlations between
$n_{R,\uparrow}$ and $n_{L,\uparrow}$, as well as $n_{R,\downarrow}$
and $n_{L,\uparrow}$, so we consider two types of correlation
functions; correlations between atoms in the same spin state
\bea
\mathcal{G}_{\uparrow \uparrow}(q,q') & = & \lav n_{\uparrow,q} n_{\uparrow,q'} \rav - \lav n_{\uparrow,q} \rav \lav n_{\uparrow,q'} \rav
\eea
and correlations between opposite spins
\bea
\mathcal{G}_{\downarrow\uparrow}(q,q') & = & \lav n_{\downarrow,q} n_{\uparrow,q'} \rav - \lav n_{\downarrow,q} \rav \lav n_{\uparrow,q'} \rav.
\eea
We again use the same convention that $q$ is located near the right
Fermi point and $q'$ located near the left Fermi point.  By
considering the dominant contraction of these correlation functions,
analogous to Eq. (\ref{contract1}) and Eq. (\ref{contract2}), we can expect the
following signatures: For CDW order we expect
$\mathcal{G}_{\uparrow\uparrow}(q,q') \sim -\delta_{q,q'}$, for SDW
$\mathcal{G}_{\downarrow\uparrow}(q,q') \sim -\delta_{q,q'}$ whereas
for triplet pairing we expect $\mathcal{G}_{\uparrow\uparrow}(q,q')
\sim \delta_{q,-q'}$, and for singlet pairing
$\mathcal{G}_{\downarrow\uparrow}(q,q') \sim \delta_{q,-q'}$.  Each of
these statements can be confirmed and quantified with a mean-field
calculation.
\subsection{\label{SFQO} Quasi-order}
We bosonize these fermionic fields according to:
\bea
\psi_{R/L, \uparrow/\downarrow}(x) & = & \frac{1}{\sqrt{2 \pi \alpha}}
e^{\pm i\Theta_{\uparrow/\downarrow}(x)+i \Phi_{\uparrow/\downarrow}(x)}
\label{bosonization_spin}
\eea
with the same definitions of $\Theta_{\uparrow/\downarrow}$ and
$\Phi_{\uparrow/\downarrow}$ that we used for spinless fermions.  We
introduce spin and charge fields according to:
\bea
\Theta_{\rho,\sigma} & = & \frac{1}{\sqrt{2}}(\Theta_\uparrow \pm \Theta_\downarrow)\\
\Phi_{\rho,\sigma} & = & \frac{1}{\sqrt{2}}(\Phi_\uparrow \pm \Phi_\downarrow)
\eea
Written in terms of $\Theta_\sigma$ and $\theta_\rho(x)=\Theta_\rho(x)
- k_f x$ the action of any SU(2)-symmetric system separates into a
charge and a spin sector:
\bea
S & = & S_\rho +S_\sigma
\eea
with:
\bea
S_\rho &= & \frac{1}{2 \pi K_\rho}\int \frac{1}{v_\rho}(\partial_\tau\theta_\rho)^2 + v_\rho (\partial_x\theta_\rho)^2
\eea
and:
\bea
S_\sigma &= & \frac{1}{2 \pi K_\sigma}\int \frac{1}{v_\sigma}(\partial_\tau\Theta_\sigma)^2 + v_\sigma (\partial_x\Theta_\sigma)^2\\
& & + \frac{2g_{1,\perp}}{(2\pi\alpha)^2}\int \cos(\sqrt{8 K_\sigma}
\Theta_\sigma)
\eea
Each of these sectors is characterized by a velocity $v_{\rho/\sigma}$
and a Luttinger parameter $K_{\rho/\sigma}$.
In addition to the quadratic terms in the action we find a non-linear
term, describing backscattering processes in the spin sector, with the
prefactor $g_{1,\perp}$.  If this action is derived from a system with
a short-ranged interaction such as (\ref{Hub}) and (\ref{MixF}), these
parameters have the following properties:
For repulsive interaction, one finds $K_\rho<1$, $K_\sigma>1$ and
positive backscattering $g_{1,\perp}$, for attractive interaction
$K_\rho>1$, $K_\sigma<1$, and negative backscattering $g_{1,\perp}<0$.
As discussed in \cite{SG, review}, this sine-Gordon model, can be
treated with an RG calculation to identify two limiting cases: The
case in which the backscattering term is irrelevant and the system
flows towards the non-interacting fixed point ($K_\sigma\rightarrow
1$), which happens for repulsive interaction, and the case in which
the backscattering term is relevant and a spin gap appears
($K_\sigma\rightarrow 0$), which happens for attractive interaction.
In the evaluation of the noise correlation functions we will use these
limiting values, $K_\sigma=1$ for the gapless phase, $K_\sigma=0$ for
the spin-gapped regime.  One can find the phase diagram of this system
in exact analogy to the spinless case,
by studying the correlation functions of the order parameters.  The
scaling exponents of these operators are given by:
\bea
\alpha_{SS} & = & 2 - K_\rho^{-1} - K_\sigma\label{a_ss}\\
\alpha_{TS} & = & 2 - K_\rho^{-1} - K_\sigma^{-1}\label{a_ts}\\
\alpha_{SDW} & = & 2 - K_\rho - K_\sigma^{-1}\label{a_sdw}\\
\alpha_{CDW} & = & 2 - K_\rho - K_\sigma.\label{a_cdw}
\eea
From these expressions one can read off the structure of the phase
diagram.  In the gapless phase we have $K_\sigma=1$, therefore singlet
and triplet pairing, as well as SDW and CDW are algebraically
degenerate.  For $K_\rho>1$ we find a TS/SS phase, for $K_\rho<1$ we
obtain a SDW/CDW phase.  For $K_\sigma\rightarrow 0$, both
$\alpha_{SDW}$ and $\alpha_{TS}$ are sent to $-\infty$, whereas
$\alpha_{SS}$ and $\alpha_{CDW}$ are now given by
$\alpha_{SS}=2-K_\rho^{-1}$ and $\alpha_{CDW}=2-K_\rho$.  Hence, we
can distinguish four regimes: For $K_\rho>2$ we have singlet pairing,
for $K_\rho<1/2$ we get CDW ordering.  In between these two values of
$K_\rho$ the system shows coexisting orders, that is, both the singlet
pairing susceptibility and the CDW susceptibility are divergent.  For
$1/2<K_\rho<1$ CDW is dominant and SS is subdominant, for $1<K_\rho<2$
it is the other way around.

\begin{figure}
\includegraphics[width=8.5cm]{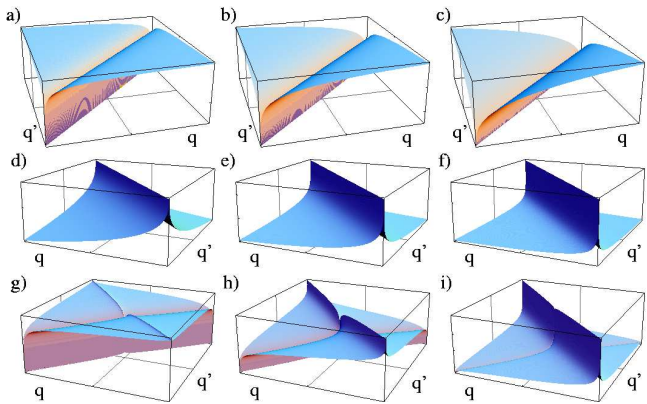}
\caption{\label{noise_S_half} Noise correlations $\lav n_{\uparrow,q}
  n_{\uparrow,q'} \rav - \lav n_{\uparrow,q} \rav \lav n_{\uparrow,q'}
  \rav$ (a -- c), $\lav n_{\downarrow,q} n_{\uparrow,q'} \rav - \lav
  n_{\downarrow,q} \rav \lav n_{\uparrow,q'} \rav$ (d -- f), and $\lav
  n_{tot,q} n_{tot,q'} \rav - \lav n_{tot,q} \rav \lav n_{tot,q'}
  \rav$ (g -- i) of a spin-$1/2$ Fermi system in 1D in the spin-gapped
  phase for different values of the Luttinger parameter
  $K_\rho$. $K_\rho=0.8$ for (a, d, g), $K_\rho=1$ for (b, e, h), and
  $K_\rho=1.25$ for (c, f, i).  In (a--c) we can see CDW ordering in
  the $\uparrow\uparrow$-channel, in (d--f) we see singlet pairing in
  the $\downarrow\uparrow$-channel. The noise correlations in the
  total density clearly show the coexistence of orders. For $K_\rho<1$
  CDW is dominant, and singlet pairing is subdominant, for $K_\rho>1$
  it is the other way around.  }
\end{figure}
The noise correlations can be calculated in the same way as described
for the spinless case.  We obtain analogous expressions to
Eq. (\ref{CorrInt}), in which the exponents $g$ and $h$ are replaced
by:
\bea
g_{\uparrow\uparrow/\downarrow\uparrow} & = & (K_\rho+K_\rho^{-1}+
 K_\sigma+K_\sigma^{-1}-4)/8
\eea
and:
\bea
h_{\uparrow\uparrow} & = & (K_\rho-K_\rho^{-1})/8+
(K_\sigma-K_\sigma^{-1})/8\label{hupup}\\
h_{\downarrow\uparrow} & = & (K_\rho-K_\rho^{-1})/8-
(K_\sigma-K_\sigma^{-1})/8\label{hdownup}
\eea
In order to understand in what regimes of the phase diagram we should
expect algebraic divergencies, we again consider the equal-time
correlation functions of the operators (\ref{op_ss})--(\ref{op_cdw}).
In momentum space, these correlation functions scale as
$|q|^{1-\alpha}$, where $\alpha$ is the corresponding scaling
exponent, given in (\ref{a_ss})--(\ref{a_cdw}).  If the system is in
the gapless phase (i.e. $K_\sigma=1$), these correlation functions
never exhibit an algebraic divergence, and we should expect to find
coexisting fluctuations throughout the phase diagram for
$g_{1,\perp}>0$, similar to the regime $1/2<K<2$ for spinless
fermions.  If the system is in the spin-gapped phase, we find the
following behavior: Both TS and SDW fluctuations are frozen out,
i.e. only short-ranged, whereas SS and CDW are increased by $1$,
compared to the gapless phase, because $K_\sigma\rightarrow 0$.  We
therefore expect algebraic divergencies for $K_\rho<1$ in the
$\uparrow\uparrow$ channel, and for $K_\rho>1$ in the
$\uparrow\downarrow$ channel.

A numerical study of the noise correlations confirms these
expectations: We indeed find coexisting fluctuations in the gapless
phase for any value of $K_\rho$.  Furthermore, since the expressions
(\ref{hupup}) and (\ref{hdownup}) become identical for $K_\sigma=1$,
we find that
$\mathcal{G}_{\uparrow\uparrow}(q,q')=\mathcal{G}_{\downarrow\uparrow}(q,q')$
in this regime. This is a manifestation of the degeneracy (at the
algebraic level) of triplet and singlet pairing for $K_\rho>1$, and of
spin density and charge density wave ordering for $K_\rho<1$, as
discussed in [\onlinecite{review}].  For the spin-gapped phase
($K_\sigma\rightarrow 0$) this symmetry is broken and
$\mathcal{G}_{\uparrow\uparrow}$ and
$\mathcal{G}_{\uparrow\downarrow}$ behave qualitatively different.  We
find the following behavior: $\mathcal{G}_{\uparrow\uparrow}(q,q')$
shows an algebraic divergence of the p-h type for $K_\rho<1$, as
expected from the equal-time correlation function of the CDW order
parameter, an algebraic cusp for $1<K_\rho<2$, and no ordering for
$K_\rho>2$.  $\mathcal{G}_{\downarrow\uparrow}(q,q')$ behaves in a
complementary way: an algebraic divergence of the p-p type is found
for $K_\rho>1$, an algebraic cusp for $1/2<K_\rho<1$ and no ordering
below that.  In particular, for $K_\rho$ in the vicinity of $1$, we
find coexisting orders, as we demonstrate in Fig. \ref{noise_S_half}.
This is particularly clear if we consider the noise correlations of
the total density $n_{tot, q} = n_{\uparrow, q} + n_{\downarrow, q}$,
for which the noise correlations are given by
$\mathcal{G}_{tot}(q,q')=2\mathcal{G}_{\uparrow\uparrow}(q,q') +
2\mathcal{G}_{\downarrow\uparrow}(q,q')$.  In Fig.  \ref{noise_S_half}
g) -- i) we clearly see the coexistence of pairing and CDW ordering.

To understand this behavior further in this limit, we use the same
argument as for the $K< 1/2$ and $K>2$ regimes for spinless fermions.
We re-write $\lav n_{\ua,q}n_{\ua,q'}\rav$ and $\lav
n_{\ua,q}n_{\da,q'}\rav$ in the same way as (\ref{other_v}), where $g$
and $h$ need to be replaced by $g_{\ua\ua/\ua\da}$ and
$h_{\ua\ua/\ua\da}$, respectively.  In the limit $K_\sigma\rightarrow
0$ the arguments that lead to the scaling behavior (\ref{scaling1})
and (\ref{scaling2}) become exact, and we obtain
\bea
\lav n_{\ua,q}n_{\ua,q'}\rav
 & \sim & \mathrm{sgn}(K_\rho-1)|q-q'|^{K_\rho-1}\\
\lav n_{\ua,q}n_{\da,q'}\rav
& \sim & \mathrm{sgn}(K_\rho^{-1}-1)|q+q'|^{K_\rho^{-1}-1}.
\eea
These expressions show exactly the structure that was found
numerically: algebraic divergencies for $K_\rho<1$ ($K_\rho>1$) in the
$\ua\ua$ ($\ua\da$) channel, and an algebraic cusp for $1<K_\rho<2$
($1/2<K_\rho<1$).

\section{\label{B}Bosons}
We turn to address the noise correlations in bosonic systems
with either long range or quasi long-range order in the off diagonal density matrix
$\langle b^\dagger(x) b(0)\rangle$. As in the fermion case we shall start with
the case of true long range order, relevant to three dimensional systems at temperature $T<T_c$. This analysis is also relevant for lower dimensional systems if they are sufficiently weakly interacting. Then the fact that the off diagonal density matrix decays as a power law is unnoticeable in a condensate of realistic size.
We shall discuss in some detail the
possibility of observing the pairing correlations associated with quantum depletion and show how such measurements would depend on the temperature.
Then we move on to derive the noise correlations in a one dimensional Bose system at zero temperature, taking into account the power-law behavior of the correlations. As in the case of Fermions we use the effective Luttinger liquid theory, which correctly accounts for the singular contributions to the noise correlations due to the long distance power-law behavior of the off diagonal density matrix.
\subsection{\label{BEC}Bose-Einstein condensate with true ODLRO}
Our starting point for analysis of the noise correlations
is the Hamiltonian of a weakly interacting Bose gas with contact interactions
\bea\label{H_bos}
H & = & \sum_\bk \epsilon_\bk a_\bk^\dagger a_\bk + \frac{U}{2 V}\sum a^\dagger_{\bk+\bq}
a^\dagger_{\bp-\bq}a_\bp a_\bk.
\eea
Here $\epsilon_\bk =
k^2/2 m$ is the free particle dispersion with $m$ and $U = 4\pi a_s/m$ with $a_s$ the
s-wave scattering length.  To compute the correlations in the condensed phase we apply the standard Bogoliubov theory (see e.g. \cite{PethickSmith}). As usual, the
operators $a\yd_0$ and $a\nd_0$ are replaced by a number representing
the condensate amplitude $\sqrt{N_0}$, while the other modes are treated as fluctuations and expanded to quadratic order. The effective Bogoliubov Hamiltonian
is then given by
\be
H_B =\sum_\bk \left(\epsilon_\bk+ U\rho_0\right) a_\bk^\dagger a_\bk
+\frac{U\rho_0}{2}\sum_\bk a\yd_{-\bk}a\yd_\bk,
\label{HB}
\ee
where $\rho_0=N_0/V$ is the condensate density.
This Hamiltonian is diagonalized by the Bogoliubov transformation
$a_\bk = u_\bk \a_\bk + v_\bk\a^\dagger_{-\bk}$ with
$u_\bk^2 = (\omega_\bk +\epsilon_\bk + U\rho_0)/2
\omega_\bk$, $v_k^2 = (-\omega_\bk + \epsilon_\bk +U\rho_0)/2\omega_\bk$,
and $\w_\bk=\sqrt{\epsilon_\bk(\epsilon_\bk +2U\rho_0)}$.

The structure of the ground
state wave function in the Bogoliubov approximation is given by
\be
\ket{\Psi_B}\sim \exp\left(\sqrt{N_0} a\yd_0+\sum_{\bk\ne 0}
(v_\bk/u_\bk)a\yd_{-\bk} a\yd_\bk\right)\ket{0}.\label{PsiB}
\ee
Like the $BCS$ state (\ref{BCSWF}), the Bogoliubov wave-function
describes perfectly correlated pairs of particles at momenta $\bk$ and $-\bk$,
which suggests the appearance of pairing correlations in the noise.
\begin{figure}
\includegraphics[width=5.5cm]{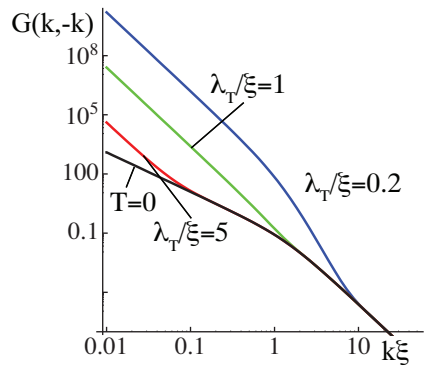}
\caption{\label{noiseloglog}
 Noise correlation of pairs $\mathcal{G}(k,-k)$ of
 a BEC,
 as a function of $k \xi$, for different ratios of $\lambda_T/\xi$.
}
\end{figure}

It is straight forward to compute the noise correlations $G(\bk,\bk')$
for $\bk,\bk'\ne 0$.
Because the Bogoliubov Hamiltonian (\ref{HB}) is quadratic we can use Wick's theorem
to decouple the four point function
\be
\mathcal{G}(\bk,\bk')=\av{n_\bk}(1+\av{n_\bk})\d_{\bk\bk'}
+|\av{a\yd_{-\bk}a\yd_{\bk}}|^2\d_{\bk,-\bk'}
\ee
where the expectation values correspond to thermal averages, and $n_\bk=a\yd_\bk a\nd_\bk$.
A bit more care is needed if either $\bk=0$ or $\bk'=0$ because the quadratic hamiltonian describes only the fluctuations in the depletion cloud, not in the condensate number. To
obtain the fluctuations in the condensate within Bogoliubov theory, we use
the conservation of total particle number, which  implies that fluctuations in the condensate number are exactly to minus those of the depletion cloud.
In other words we substitute
$n_0=N-\sum_{\bk\ne 0} n_\bk$ for the condensate particle number {\em operator}.
Then we may use (\ref{HB})
to compute the noise correlation between points $k=0$ and $k'$.
Putting it all together we get the general expression for the noise correlations:
\bea
\mathcal{G}(\bk,\bk')&=& g_\bk\d_{\bk,-\bk'}+f_\bk\d_{\bk\bk'}
-h_\bk(\d_{\bk 0}+\d_{\bk'0})\nn\\
&&+\d_{\bk 0}\d_{\bk'0}\sum_\bq h_\bq
\label{Gbose}
\eea
where
\bea
g_\bk&=& u_\bk^2 v_\bk^2\left(1+ 2 \av{n_{\a,\bk}}\right)^2 \nn\\
f_\bk&=& g_\bk+ \av{n_{\a,\bk}}\left(1+\av{n_{\a,\bk}}\right) \nn\\
h_\bk&=& 2g_\bk+ \av{n_{\a,\bk}}\left(1+\av{n_{\a,\bk}}\right)
\label{Gdetails}
\eea
Here $\av{n_{\a,\bk}}=\av{\a\yd_\bk\a\nd_\bk}=[\exp(\w_\bk/T)-1]^{-1}$
is the quasi-particle number distribution.

At zero temperature each term in (\ref{Gbose}) has a simple physical interpretation.
We already noted that the first term manifests the pairing correlations
present in the quantum depletion described by the
Bogoliubov wave function (\ref{PsiB}). The second term is a positive correlation
due to boson bunching at a point in $\bk$-space. The {\em dips} at $\mathcal{G}(\bk,0)$ reflect the fact that an extra atom found at $k$ in the quantum depletion cloud corresponds to a pair of atoms,
now {\em missing} from the condensate. Finally the positive peak at $\mathcal{G}(0,0)$
appears because extra atoms in the condensate must always come in pairs, annihilated from the depletion cloud. That is, if we find an extra atom in the condensate, then we are sure to find another extra atom in it.

The evolution of the peaks with temperature and their momentum dependencies are controlled by the ratio of two natural length scales of the problem.
One is the healing length of the condensate which is determined
by interactions $\xi=1/\sqrt{2mU\rho_0}$.
The other is the thermal wavelength $\lambda_T=1/\sqrt{2 m
  T}$. The combination $u_\bk^2 v_\bk^2$ is of course temperature independent and may be expressed using the healing length alone as $[8(\xi  k)^2+4(\xi k)^4]^{-1}$.
  The temperature dependence arises from the ratio $\w_\bk/T$, which appears in the
  distribution function, and may be expressed as
$(\lambda_T/\xi)^2 (k\xi)\sqrt{1 + (k\xi)^2}$. Note that the dimensionless ratio
$(\lambda_T/\xi)^2$ is the same as the ratio $\mu/T$.

An interesting correlation to observe is the pairing correlation
at $\bk'=-\bk$, which can be written explicitly using (\ref{Gbose}) and (\ref{Gdetails}) as
\be
\mathcal{G}(\bk,-\bk)={\coth^{2}\left[(\mu/T) (k\xi)\sqrt{1 + (k\xi)^2}\right]\over 8(\xi  k)^2+4(\xi k)^4}
\ee
This function is plotted on a log-log scale in
Fig. \ref{noiseloglog}. The behavior of the peak weight at small relative
momentum $k\xi<<1$ has a very simple form. First, at very low temperatures,
such that $(\mu/T)k\xi >>1$ we have $\mathcal{G}(\bk,-\bk)\approx 1/[8(k\xi)^2]$.
This is the power law seen for $T=0$ in Fig. \ref{noiseloglog}. At higher temperature
or sufficiently small momentum such that $(\mu/T)k\xi <<1$ we have
$\mathcal{G}(\bk,-\bk)\approx (T/\mu)^2/[2(k\xi)^4]$, which is seen in the other curves in the same figure.

\begin{figure}
\includegraphics[width=5.5cm]{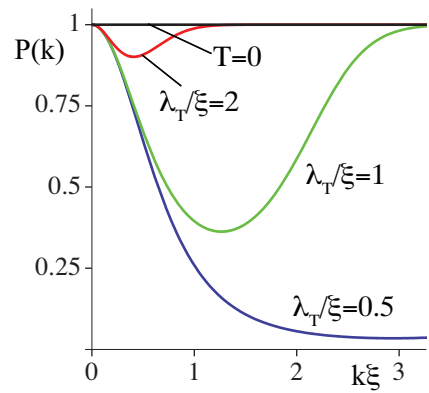}
\caption{\label{pairnoise}
Pair correlation function $P(k)=\mathcal{G}(k,-k)/\mathcal{G}(k,k)$,
 as function of $k \xi$, for different ratios of $\lambda_T/\xi$.
}
\end{figure}

It is interesting to note that the pairing correlations are substantially
enhanced with temperature. This seems surprising given that the
origin of the pairing is the quantum depletion
in the ground state (\ref{PsiB}). The effect may be interpreted as Bose enhancement of paired thermal fluctuations. However we should also note that the overall noise level is also growing with temperature, that is the local (in $\bk$) particle number fluctuation $G(\bk,\bk)$ is increasing even more steeply with $T$.
In Fig. \ref{pairnoise} we plot the pair correlation normalized by the local number
fluctuation
\be
P(\bk) =  \frac{\mathcal{G}(\bk,-\bk)}{\mathcal{G}(\bk,\bk)}
\ee
Clearly, for $T=0$  (i.e. $\lambda_T/\xi \rightarrow \infty$) we have $P(k)=1$
Because in the ground state (\ref{PsiB}) the number fluctuations always come in opposite momentum pairs the particle number at $k$ and $-k$ must be identical, and hence
also the correlations $\mathcal{G}(k,-k)$ and $\mathcal{G}(k,k)$ are equal.
At finite temperature the pairing correlations are suppressed compared to
the local (in $\bk$) number fluctuation. The ratio remains 1 at very high momentum because at $\w_\bk>>T$ thermal occupation (which is exponentially suppressed is negligible compared to the quantum depletion. More interesting is the fact that
$P(\bk)$ approaches $1$ also in the limit of small relative momenta, which may again
be a signature of Bose enhancement of the pairing fluctuations.

\subsection{\label{QC}Quasi-condensate}
In this section we discuss the noise correlations for a LL of bosons.
We use Haldane's representation \cite{Haldane} of a bosonic operator,
defined as:
\bea
b(x) & = & [\rho_0 + \Pi(x)]^{1/2} \sum_m e^{2im\Theta(x)} e^{i\Phi(x)}
\eea
The fields $\Theta(x)$, $\Phi(x)$ and $\Pi(x)$ are defined in the same
way as for fermionic LLs.  $\rho_0$ is the average density.  Note that
now the sum is over the even harmonics, $2m$, and not the odd ones,
$2m+1$, that are used to represent a fermionic operator.  The action
of the system can be written as:
\bea
S & \sim &
\frac{1}{2 \pi K}\int d^2x \partial_\mu\theta \partial^\mu \theta \sim  \frac{K}{2 \pi}\int d^2x \partial_\mu\Phi\partial^\mu\Phi
\eea
As for fermionic systems, these representations
 are just quadratic in the fields, therefore all
 correlation functions reduce to Gaussian integrals.

 For small momenta $k$ and $k'$ the noise correlations are
\bea \label{noise_B}
\mathcal{G}(k,k') & \sim &  \rho_0^2 \int e^{i k x_{12} + i k x_{34}}
\mathcal{F}(x_{12}) \mathcal{F}(x_{34}) (\mathcal{A}_h -1),
\eea
 which we derive in detail in App. \ref{NoiseBose}.
 $\mathcal{F}$ and $\mathcal{A}$ are of similar form as before,
 with $g = 1/(4K)$ and $h=-1/(4K)$.

\begin{figure}
\includegraphics[width=8cm]{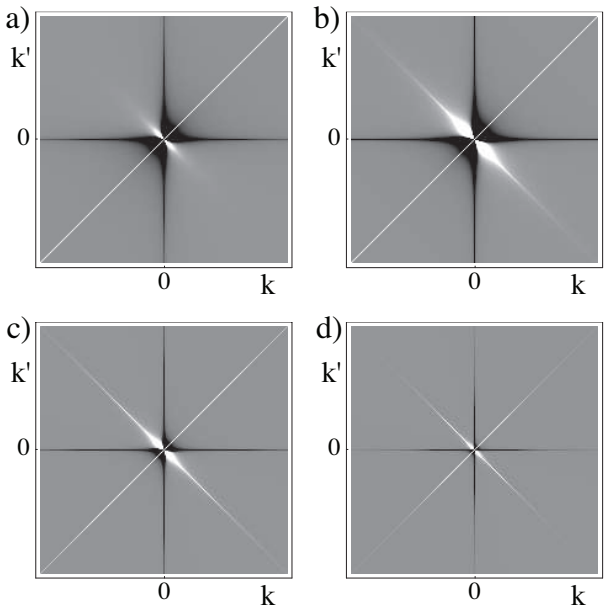}
\caption{\label{noise_Bos} Noise correlations $\mathcal{G}(k,k')$ of a
  1D Bose system for $K=1.05$, $2.5$, $8$ and $20$ (a--d).  The grey scale is
  linear, with light colors corresponding to positive values, and dark
  colors indicating negative values.  The center of the plot
  corresponds to $(0,0)$.  For large values of $K$, the properties of
  a 1D bosonic superfluid approach the ones of a BEC with various
  sharply peaked structures.  For smaller values of $K$, the pairing
  peak and the quasi-condensate dips broaden and become weaker.  }
\end{figure}
This expression can be numerically evaluated for a finite system,
 with the replacement $x\rightarrow L \sin(2 \pi x/L)/2\pi$.  In
Fig. \ref{noise_Bos} we show $\mathcal{G}(k,k')$, plotted for
different values of $K$.

For large values of $K$, the LL results resemble qualitatively the BEC
result: The noise correlation function shows both a sharp pairing and
bunching contribution which are equal in magnitude, as well as sharp
quasi-condensate contributions along $(k,0)$ and $(0,k')$.  In
addition there is a large peak at ($k=0$,$k'=0$), (invisible in the
representation in Fig. \ref{noise_Bos}).  As we reduce $K$ to smaller
values (corresponding to larger repulsive interactions between the
bosons), the quasi-condensate gets visibly broadened. This is due to
the fact that there is no true condensate in 1D: The occupation in
$k$-space is not peaked as a $\delta$-function, but only an algebraic
divergence, which gets broader for smaller values of $K$.  As a
consequence, also the pairing peak gets broadened, because the
distribution of the total momentum of the pairs that are created from
the condensate gets broadened.  Furthermore, the overlap with the
broad quasi-condensate dips diminishes the magnitude of the pairing
peaks for smaller values of K, and eventually, for $K=1$, which is the
Tonks-Girardeau limit, the pairing peak is entirely surpressed.

In contrast to the broadening of the pairing peak and the
quasi-condensate dips, the bunching peak is a $\delta$-function for
all values of $K$.  This arises because we have $|g|=|h|$, and
therefore the integrand in Eq. (\ref{noise_B}) does not fall off in
one direction.  This can be understood from the integral expression
(\ref{noise_B}).  We introduce the variables $x=(x_{12}-x_{34})/2$,
$z=(x_{12}+x_{34})/2$ and $y=x_{23}+z$, and rearrange the integral
(\ref{nnBos}):
$\lav n_{k}n_{k'} \rav  =  \int e^{i\Delta k x + i \bar{k}z}
\mathcal{F}(z+y)\mathcal{F}(z-y) \tilde{A}$.
 Here we introduced the definitions $\Delta k =k-k'$, $\bar{k}=k+k'$, and
\bea\label{Atildeh}
\tilde{A} & = & \Big(\frac{(x_0^2 + (x+z)^2)(x_0^2 + (x-z)^2)}{(\
x_0^2 + (x+y)^2)(x_0^2 + (x-y)^2)}\Big)^h.
\eea
From this, it is clear that for $x\rightarrow \infty$ the integrand
approaches $1$, and does not fall off to zero, giving rise to a
$\delta$-function.

Finally, we discuss the transition of a 1D bosonic superfluid to a
Mott insulator, which can occur if there is a lattice potential
present that is commensurate to the density of the superfluid.  For
$k,k'\approx 0$ and for $K\rightarrow 0$, $\tilde{A}$ approaches 1, as
can be seen from (\ref{Atildeh}).  In this limiting case, the integral
of $x$ becomes $\delta_{\Delta k}=\delta_{k,k'}$.  The remainder of
the integral can be evaluated to be $\lav n_k\rav^2$ for $k=k'$, so
the entire singular contribution is given by $\lav n_k\rav ^2
\delta_{k,k'}$.  Therefore, for $K\rightarrow 0$, which describes the
Mott insulator transition, we find $\tilde{\mathcal{G}}(k,k')
\rightarrow \lav n_k\rav^2 \delta_{k k'} - \lav n_k\rav\lav
n_{k'}\rav$, as for the higher dimensional case \cite{ehud}.  By using
the higher modes of Haldane's representation, we can determine the
behavior of $\mathcal{G}(k,k')$ for $k\approx 2nk_b$ and $k'\approx
2mk_b$.  We use the expression $b \sim \sqrt{\rho_0}e^{2 i n
  \Theta}e^{i\Phi}$, and we get: $\lav n_k\rav = \rho_0 \int dx_{12}
e^{i k x_{12}} \mathcal{F}_{n}(x_{12})$, with $g$ given by $g=1/4K+n^2
K$.  For $\lav n_k n_{k'}\rav$ we obtain $\lav n_k n_{k'}\rav =
\rho_0^2 \int e^{i k x_{12} + i k x_{34}} \mathcal{F}_{n}(x_{12})
\mathcal{F}_{m}(x_{34}) \mathcal{A}$.  $\mathcal{A}$ is defined as
before, with $h$ now given by $h=-nmK-1/4K$.  We can now go through
the same steps that were used to identify the $\delta$-function for
$k,k'\approx 0$. We find that for any $n$ and $m$, the noise
correlation function approaches the peaked shape that was found for
$k,k'\approx 0$ in the limit $K\rightarrow 0$.  This coincides with
the result found in \cite{ehud}, in which an ansatz of the form $| MI
\rav = \prod_i b_i^{\dagger} | 0 \rav$ was used to derive this result.
\section{\label{Conclusion}Conclusions}
We have investigated the nature of noise correlations in systems of ultra cold atoms, that do not necessarily support long range order in any order parameter.
The essential difference between noise correlations (\ref{G}), which may be measured in time of flight experiments\cite{BlochHBT,PortoNoise,BlochFermions,markus}, and standard order parameter correlations, is that the noise correlations are non local in real space. In other words, they are not a fourier transform of a two point correlation function in real space. In this paper we investigated, the general connections, that nevertheless exist between these two types of correlations in many-body systems. We then derived the singular contributions to the noise correlation function for several ultra-cold boson and fermion systems of interest. Our focus was on low dimensional systems, which display power law order-parameter correlations.

  For Fermi systems, we showed in sec \ref{subsec:LRO}), that true long range order in a particular order parameter (spin,density or pairing) leads to a delta-function contribution to the noise correlations. It is tempting to assume that if there are power-law decaying correlations in the same order parameter, then they would contribute corresponding algebraic singularities to the noise. However, we showed that this is only the case provided the algebraic decay is sufficiently slow. For example in a system of one dimensional spinless Fermions, the noise correlations appear similar to the mean field result, albeit with power-law peaks, only for $K< 1$ (strong repulsive interaction) or $K>2$ (strong attractive interaction). In the intermediate regime $2>K>1/2$, they display a unique structure that reveals both the dominant and sub-dominant order parameter correlations.
  
In a system of spinfull fermions, on the other hand, the noise correlations display significant singular contributions only in the spin-gapped phases (i.e. for negative backscattering). In this regime they provide information on both the singlet-pairing and density wave correlations.
  
  In all cases we showed, that the noise correlations treat the particle-particle (pairing) and particle-hole (spin or density) channels on the same footing. Noise correlations are thus sensitive, and can easily distinguish between the dual order parameters. In general, pairing correlations manifest as positive noise  correlations (i.e. peaks), while spin and charge correlations appear as anti-correlations in the noise (dips). This is in marked contrast to external probes, such as Bragg scattering, which couple only to spin or particle densities (particle-hole channel).

 When considering Bose systems in section \ref{B}, we first treated a condensate with true off diagonal long range order using Bogoliubov theory. The most interesting feature in this case is a correlation peak $\sim\d(\bk+\bk')$, which is a direct manifestation of the pairing correlations in the quantum depletion cloud. We find, somewhat counter intuitively, that these correlations are enhanced with increasing temperature at low temperatures. It should be noted however that the "normalized" pairing correlation, that is, relative to the number fluctuation on the $k$-point, is indeed suppressed with temperature. Additional peaks appear due to correlations between the condensate number and the number of particles in the depletion cloud. These appear on the lines $k=0$ and $k'=0$.

 Finally we used the quantum hydrodynamic (Luttinger liquid)description \cite{Haldane} of the Bose liquid, to address one dimensional Bose systems at $T=0$. For this system, characterized by algebraic decay of off diagonal order, the pairing correlation in the noise also converts to a power-law singularity.

\section{Acknowledgments}
We thank E. Demler for stimulating discussions. This work was supported by US israel binational science foundation (E. A.), and the Israeli Science foundation (E. A.).

\appendix
\section{}\label{noisederiv}
In this appendix we derive Eq. \ref{CorrInt}, using bosonization.  We
write $\lav n_{q} n_{q'} \rav$ and $\lav n_{q} \rav$ as
\bea
\lav n_{q}n_{q'}\rav & = &  \frac{1}{L^2}\int \prod_{i=1}^{4}
dx_i e^{i q x_{12} +
i q' x_{34}}\nonumber\\
& & \lav\psi_{R}^\dagger (1) \psi_{R}(2) \psi_{L}^\dagger (3)
\psi_{L}(4)\rav\label{nRnL_1}\\
\lav n_{q}\rav & = & \frac{1}{L} \int d1 d2 e^{i q x_{12}}
\lav \psi_{R}^\dagger (1) \psi_{R}(2)\rav.\label{n_1}
\eea
$q$ is the momentum relative to the right Fermi point, $q=k-k_F$, and
$q'$ the one relative to the left Fermi point, $q'=k'+k_F$, as before.
We use the notation '1' for '$x_1$', etc., and $x_{12}$ for $x_1-x_2$,
etc.  An analogous expression holds for $\lav n_{q'}\rav$, with `R'
replaced by `L'.

Using the bosonization expression in Eq. (\ref{bosonization}) for
$\lav n_q \rav$ we obtain
\bea
\lav n_{q}\rav& = & \int\frac{d1 d2}{2\pi\alpha L} e^{i q x_{12}}
\lav e^{-i \Theta(1)-i \Phi(1)} e^{i\Theta(2)+i\Phi(2)}\rav
\eea
and a similar expression for $n_{q'}$ with $\Theta(2)\rightarrow
-\Theta(2)$ and $-\Theta(1)\rightarrow\Theta(1)$.
By using $e^{A}e^B = e^{A+B} e^{[A,B]/2}$ we get:
\bea
\lav n_{q} \rav& = & \int\frac{d1 d2}{2\pi\alpha L} e^{i q x_{12}}
\lav e^{i (\Theta(2)-\Theta(1))+i (\Phi(2)-\Phi(1))}\rav\nonumber\\
& & e^{[\Theta(1),\Phi(2)]/2 + [\Phi(1),\Theta(2)]/2}
\eea
The commutator between the fields $\Theta(x)$ and $\Phi(x)$ is in
Eq. \ref{com}.  Next we use $\lav e^A\rav=e^{\lav A^2\rav/2}$.  Here
it is necessary to impose a short distance cut-off $x_0$ on the
interaction, that is, $K$ needs to depend on the momentum and has to
fall off exponentially to 1 for momenta of the order of $1/x_0$.  The
expression that we use is
\bea
\lav (\Theta (x) - \Theta (0))^2 \rav & = & \frac{K-1}{2} \log\frac{x_0^2 + x^2}{x_0^2}  +\frac{1}{2} \log\frac{\alpha^2 + x^2}{\alpha^2}
\nonumber\eea
An analogous expression holds for $\Phi(x)$ with $K$ replaced by
$K^{-1}$.  The cut-off $x_0$ on the interactions will stay finite
throughout the calculation and can be interpreted as an effective
bandwidth.  If we apply this to $n_{q (q')}$, we obtain
\bea
\lav n_{q (q')}\rav& = & \frac{1}{2 \pi} \int dx_{12} e^{i q x_{12}} \mathcal{F}^{(*)}(x_{12})
\eea
where we introduced
\bea
\mathcal{F}(x) & \equiv & \Big(\frac{x_0^2}{x_0^2+x^2}\Big)^g\frac{1}{\alpha - i x}
\eea
with $g$ given by $g=(K+K^{-1}-2)/4$.

We can evaluate $\lav n_{q} n_{q'} \rav$ along the same lines.  We
again use Eq. \ref{bosonization}, rearrange the exponents in the same
way as we did for $n_{q}$, while keeping track of the non-vanishing
commutators between them, and take the expectation value, to obtain
\bea
\lav n_{q}n_{q'}\rav & = & \int\frac{e^{i q x_{12} +
i q' x_{34}}}{(2\pi\alpha)^2 L^2}
e^{-\lav(\Theta(1)-\Theta(2)-\Theta(3)+\Theta(4))^2\rav/2}\nonumber\\
& & e^{-\lav(\Phi(1)-\Phi(2)+\Phi(3)-\Phi(4))^2\rav/2}\nonumber\\
& & e^{[\Theta(1),\Phi(2)]/2 + [\Phi(1),\Theta(2)]/2}\nonumber\\
& & e^{-[\Theta(3),\Phi(4)]/2 - [\Phi(3),\Theta(4)]/2}
\eea
This can be evaluated to
\bea
\lav n_{q}n_{q'}\rav & = & \int\frac{e^{iqx_{12}+iq'x_{34}}}{(2\pi)^2 L} \mathcal{F}(x_{12})\mathcal{F}^*(x_{34})\mathcal{A}
\eea
The integration in this expression is over the three spatial variables
$x_{12}$, $x_{23}$ and $x_{34}$.  $\mathcal{A}$ is defined as
\bea\label{Ah}
\mathcal{A} & = & \Big(\frac{(x_0^2+x_{14}^2)(x_0^2+x_{23}^2)}{(x_0^2 + x_{13}^2)(x_0^2 + x_{24}^2)}\Big)^h.
\eea
The exponent $h$ is given by $h=(K-K^{-1})/4$.  Combining the
expressions that we derived for $\lav n_{q}\rav$ and $\lav n_{q}
n_{q'}\rav$ we get for $\mathcal{G}(q,q')$:
\bea
\mathcal{G}(q, q') & = & \int\frac{e^{iqx_{12}+iq'x_{34}}}{(2\pi)^2 L} \mathcal{F}(x_{12})\mathcal{F}^*(x_{34})(\mathcal{A} -1)\nonumber
\eea
which is Eq. \ref{CorrInt}.

\section{}\label{intermediate}
In this appendix we expand $\mathcal{G}(q, q')$ to second order in the
exponent $h$. We show that the first order term vanishes, and the
second order term gives Eq. \ref{NC_int}.
The first order term is given by
\bea\label{Ah_1}
& & h \int dx_{23}\log\Big(\frac{(x_0^2+s(x_{14})^2)(x_0^2+s(x_{23})^2)}{(x_0^2+s(x_{13})^2)(x_0^2+s(x_{24})^2)}\Big)
\eea
with $s(x)$ defined as $s(x)=L/2\pi\sin(2\pi x/L)$.  Here we
explicitly kept the expression for a finite-size system. Since the
integral separates into the sum of four terms of the form $\pm \int
\log(x_0^2+s(x)^2)$, and each of these four terms integrates to the
same value (as can be seen by shifting the integration variable, and
using the periodicity of $\sin(2\pi x/L)$), the entire expression
integrates to zero.
The second order term is given by
\bea\label{Ah_2}
& & \frac{h^2}{2} \int dx_{23}\Big[\log\Big(\frac{(x_0^2+x_{14}^2)(x_0^2+x_{23}^2)}{(x_0^2+x_{13}^2)(x_0^2+x_{24}^2)}\Big)\Big]^2,
\eea
where we left out the cut-off and the finite-size representation for
notational convenience.  It can be checked that these will ensure the
following expressions and manipulations to be well-defined.

Expression (\ref{Ah_2}) corresponds to a sum of integrals of two
types: $\int dx\log^2(x^2)$ and $f(r)\equiv\int dx\log
x^2\log(x+r)^2$, as can be seen by expanding the square in
(\ref{Ah_2}).  The first integral merely provides terms that cancel
divergent terms of the second integral type in the limit $L\rightarrow
\infty$.  The second term is given by
$f(r)  =  C_1 + C_2 |r|$,
with $C_1$ and $C_2$ some constants.  This can easily be shown by
taking derivatives, and by observing that $f(r)$ is even.

With this expression for the various integrals of the type $f(r)$,
that we get from expanding the square in (\ref{Ah_2}), we get for
$\mathcal{G}(q,q')$:
\bea
\mathcal{G}(q,q')& \sim & h^2 \int \frac{e^{i q x_{12}}}{x_{12}}
 \frac{e^{i q' x_{34}}}{x_{34}} (|x_{12}+x_{34}|+|x_{12}-x_{34}| \nonumber\\
 & & -2|x_{12}|-2|x_{34}|)
\eea
To evaluate this integral we divide the integration range into eight
sectors. As an example we treat the case given by: $x_{12}>0$,
$x_{34}>0$, $x_{12}-x_{34}>0$.  (The other sectors are characterized
by similar sets of inequalities.)
For this part of the integration range we get an expression of the form:
\bea
&& -2 \int_0^\infty dx_{12} \frac{e^{i q x_{12}}}{x_{12}} \frac{1}{i q'} (e^{i q' x_{12}} -1)
\eea
The other sectors of the integration range give similar
expressions. These can be grouped into a sum of integrals that contain
integrals of the form $\int dx/x \exp(ipx)= i\pi \mathrm{sgn}(p)$,
with different combinations of $q$ and $q'$ in the exponent.  With
that we get for $\mathcal{G}(q,q')$ an expression which can be written
as
\bea
\mathcal{G}(q,q') & \sim & -h^2\mathrm{sgn}(q)\mathrm{sgn}(q')\min\Big(\frac{1}{|q|},\frac{1}{|q'|}\Big),
\eea
 which is Eq. \ref{NC_int}.

\section{}\label{NoiseBose}
We calculate $\langle n_k\rangle$ for $k\approx 0$ mode, for which the
Bose operator is given by $b\sim \sqrt{\rho_0} e^{i\Phi}$.  For $\lav
n_k \rav$ we find:
\bea
\lav n_k \rav & \sim & \rho_0 \int dx_{12} e^{ikx_{12}} e^{- \frac{1}{2}
\lav(\Phi(2)-\Phi(1))^2\rav}
\eea
For the evaluation of the expectation value
$\lav(\Phi(2)-\Phi(1))^2\rav$ we use a slightly different cut-off
procedure than for the fermions, in particular:
\bea
\lav(\Phi(2)-\Phi(1))^2\rav & = & \frac{1}{2K} \log\frac{x_0^2 + x_{12}^2}{x_0^2}.
\eea
With that we find
$\lav n_k \rav  \sim  \rho_0 \int dx_{12} e^{ikx_{12}} \mathcal{F}(x_{12})$.
where we defined:
\bea
\mathcal{F}(x) & = & \Big(\frac{x_0^2}{x_0^2 + x^2}\Big)^g
\eea
The exponent $g$ is given by $g=1/4K$.  Next we evaluate the
expectation value $\lav n_k n_{k'}\rav$ along the same lines.  We
obtain:
\bea \label{nnBos}
\lav n_k n_{k'}\rav & \sim & \rho_0^2 \int e^{i k x_{12} + i k x_{34}}
\mathcal{F}(x_{12}) \mathcal{F}(x_{34}) \mathcal{A}
\eea
$\mathcal{A}$ is defined in the same as in Eq. (\ref{Ah}), but $h$
is now given by $h=-1/4K$.  We combine these expressions to get the
correlation function $\mathcal{G}(k,k')$:
\bea \label{noise_B}
\mathcal{G}(k,k') & \sim &  \rho_0^2 \int e^{i k x_{12} + i k x_{34}}
\mathcal{F}(x_{12}) \mathcal{F}(x_{34}) (\mathcal{A} -1).
\eea
%
%


\end{document}